\documentclass[twocolumn,prb,floatfix,superscriptaddress,nobibnotes]{revtex4-2}
\usepackage[unicode=true, colorlinks=true, citecolor={blue!80!black}, urlcolor={blue!50!black}, linkcolor = {blue!80!black}]{hyperref}
\usepackage{graphicx}
\usepackage{bm}
\usepackage{array}
\usepackage{amsmath}
\usepackage{amssymb}
\usepackage{contour}

\usepackage{physics}
\usepackage[normalem]{ulem}
\usepackage{xcolor}

\usepackage{nicematrix}

\usepackage{cleveref}
\usepackage{orcidlink}

\begin{document}

\title{Nonreciprocal conductance in uniformly dissipative devices}

\author{Oliver Solow\orcidlink{0009-0004-0137-8838}}
\affiliation{Center for Quantum Devices, Niels Bohr Institute, University of Copenhagen, DK-2100 Copenhagen, Denmark}
\author{Emil J. Bergholtz\orcidlink{0000-0002-9739-2930}}
\affiliation{Department of Physics, Stockholm University, AlbaNova University Center, 106 91 Stockholm, Sweden}
\author{Karsten Flensberg}
\affiliation{Center for Quantum Devices, Niels Bohr Institute, University of Copenhagen, DK-2100 Copenhagen, Denmark}

\begin{abstract}
    When studying non-Hermitian electronic systems, an obvious question is how various non-Hermitian effects affect measurable quantities like the conductance. Here, we show that uniformly dissipative circuits can exhibit nonreciprocal conductance, meaning that the two nonlocal conductances are different. We describe how this happens through a difference in transmission times between left-moving and right-moving electrons. We consider a specific case of a dissipative Rashba nanowire with a skewed magnetic field, and show how this difference in transmission times comes about through interference inside the circuit, and how this is modified as the dissipation strength changes.
\end{abstract}
\maketitle

\section{Introduction}

Recently, there has been a growing interest in non-Hermitian physics\cite{Ashida:NonHermitianPhysics:20}, and in particular in its application to mesoscopic devices. An example of this is the non-Hermitian skin effect (NHSE), where the eigenstates of an effective non-Hermitian Hamiltonian bunch up on one side of the system \cite{Lee:AnomalousEdge:16,Xiong:Whydoes:18,Yao:EdgeStates:18,Kunst:BiorthogonalBulkBoundary:18,Okuma:TopologicalOrigin:20,Zhang:CorrespondenceWinding:20,Gohsrich:NonHermitianskin:25}.  Several recent studies have considered how to engineer and measure devices which exhibit NHSE, with ideas including spin-orbit coupled wires with ferromagnetic leads\cite{Kokhanchik:NonHermitianskin:23,Geng:Nonreciprocalcharge:23,Paya:NonHermitianskin:26} and Hall devices\cite{Ochkan:NonHermitiantopology:24,Chaturvedi:NonHermitiantopology:26}. In all of these devices, the presence of nonreciprocal transport, where the nonlocal conductances are different, is interpreted as a signature of NHSE.

For a Hamiltonian to exhibit NHSE, it must be non-normal, i.e. its Hermitian and antihermitian parts must not commute\cite{Bergholtz:Exceptionaltopology:21}. However, normal non-Hermitian Hamiltonians can still be interesting. An example of this is using Hamiltonians with a constant imaginary part to study uniform dephasing\cite{Brouwer:Voltageprobeimaginarypotential:97}, such as the Dynes parameter for disordered superconductors\cite{Dynes:DirectMeasurement:78,Eberlein:FermiSurface:16,Kontani:Intrinsicanomalous:07,Mitscherling:Longitudinalanomalous:20,Verret:Phenomenologicaltheories:17}. While these Hamiltonians, which we will dub trivially non-Hermitian Hamiltonians, have trivial spectra, it turns out that they can also exhibit non-trivial and nonreciprocal transport properties. 

In this paper, we consider the nonreciprocal transport of trivially non-Hermitian Hamiltonians and see that when transmission times in opposite directions differ, even uniform dissipation can lead to non-reciprocal transport. We show that this behavior is linked to the difference in transmission time for different directions, and show how this is related to the symmetries of the Hamiltonian. While similar ideas have been considered before\cite{Giovannelli:PhysicalInterpretation:25,Chen:Usetransmission:22,Chen:Asymmetrictransmission:24a,Shaibe:SuperuniversalStatistics:25}, we show how this effect arises from relatively simple mathematical considerations in generic systems.

We then consider the case of a Rashba nanowire with a magnetic field skewed from the spin-orbit axis. We see that nonreciprocity is present at spectral anticrossings, and explain this by considering how interference effects between the different states lead to a transmission time difference. Finally, we show how these interference effects also qualitatively describe how the nonreciprocity changes with the dissipation strength.

\subsection{Nonreciprocity and nonlinearity}

We define a nonreciprocal response to mean that the left-right differential conductance $G_{LR}=\frac{dI_L}{dV_R}$ is different from the right-left differential conductance $G_{RL}=\frac{dI_R}{dV_L}$. The nonlocal conductances thus measure the change in current in one lead in response to a voltage change in another.

To quantify the nonreciprocity, we introduce the conductance difference
\begin{equation}
    \Delta G=G_{LR}-G_{RL}.
\end{equation}
Since the nonlocal conductances measure how much the current in one lead changes due to an infinitesimal change in the voltage of another lead, the conductance difference measures the nonreciprocity of this response: If $\Delta G$ is positive in a system, voltage changes on the right lead affect the left lead strongly, whereas changes on the left lead only have a weak effect on the right lead. This picture is flipped for negative $\Delta G$.

It is worth clarifying that this nonreciprocity is unrelated to nonlinear responses\footnote{Some papers, e.g \cite{Zou:Nonreciprocalballistic:24, Anan:Nonreciprocalcurrent:26}, use ''nonreciprocity'' to indicate a nonlinear effect. For our purposes, it is important to distinguish these two effects.}. To see this clearly, we can consider a two-terminal device. In this device, charge conservation means that $I_L=-I_R=I$, while gauge invariance allows us to choose a gauge such that $V_L=-V_R=V/2$. In this case, we see that $G_{LR}=G_{RL}=G$, and as such the system is reciprocal. However, it is still perfectly possible for the current at $V$ and $-V$ to have different magnitudes:
\begin{equation}
    I(V)=\int_{0}^V dv\, G(v)\neq \int_{-V}^{0} dv\, G(v)=-I(-V),
\end{equation}
which shows that nonreciprocity and nonlinearity are two unrelated phenomena. The nonreciprocity is an effect of how the particle lifetimes differ for particles moving in different directions. It should be noted that a non-Hermitian self energy can induce interesting nonlinear effects, which has been explored in systems with impurities \cite{Michishita:Dissipationgeometry:22,Anan:Nonreciprocalcurrent:26,Tsirkin:separationHall:22}. These effects can be present in two-terminal systems but are, however, reciprocal in the sense that we define here.

Another take on this is to start with the $S$-matrix for a two-terminal setup
\begin{equation}
    \begin{pmatrix}
        r_{LL}(\omega) & t_{LR}(\omega) \\ t_{RL}(\omega) & r_{RR}(\omega)
    \end{pmatrix}.
\end{equation}
Here we, for simplicity, consider a single-band model. From  this $S$-matrix, we can calculate the nonlocal conductance by the Landauer-Büttiker formula\cite{Haug:QuantumKinetics:08}
\begin{equation}
    G_{\alpha\beta}=\frac{e^2}{h}T_{\alpha\beta}=\frac{e^2}{h}\abs{t_{\alpha\beta}}^2,
\end{equation}
where $T_{\alpha\beta}$ is the transmission probability from lead $\beta$ to lead $\alpha$. As we are considering a two-terminal system, the $S$-matrix makes it very easy to see that $\Delta G=0$ in this case: Since the $S$-matrix is unitary, $\abs{t_{LR}}=\abs{t_{RL}}$, the conductances must be the same. Below, we see how multiple leads which makes the $S$-matrix nonunitary changes this conclusion. 

\section{Conductance asymmetry}

For a multi-terminal system, the $S$-submatrix describing two of the terminals is not unitary, and hence the symmetry of the two transmission amplitudes can be broken. In previous studies, this symmetry was broken by having the additional leads induce a non-trivial non-Hermitian self energy which causes the non-Hermitian skin effect \cite{Kokhanchik:NonHermitianskin:23,Geng:Nonreciprocalcharge:23,Paya:NonHermitianskin:26}. Here, however, we will presume that the third lead is completely featureless and thus cannot cause any changes to the eigenstates. In order to address the nonreciprocity in this case, we relate the $S$-matrix to the two-terminal $S$-matrix in absence of the third lead, where we can then write $t_{\alpha\beta}=a(\omega)e^{i\phi_{\alpha\beta}(\omega)}$ where $a(\omega)$ is real.  We can now introduce the third lead by analytical continuation of the frequency as $\omega\rightarrow \omega+i\gamma$. We then obtain
\begin{equation}
\begin{split}
    T_{\alpha\beta}(\omega+i\gamma)=\abs{t_{\alpha \beta}(\omega+i\gamma)}^2\\=\abs{a(\omega+i\gamma)}^2e^{-2\Im{\phi_{\alpha\beta}(\omega+i\gamma)}}.
\end{split}
\end{equation}
Since $t_{\alpha\beta}(\omega)$ is an analytical function, we can perform a series expansion of $\Im{\phi_{\alpha\beta}(\omega+i\gamma)}$ around $\gamma=0$:
\begin{equation}\label{dphidw}
\begin{split}
    \Im{\phi_{\alpha\beta}(\omega+i\gamma)}\approx \Im{\phi_{\alpha\beta}(\omega)}+\gamma\Im{\frac{\partial\phi_{\alpha\beta}}{\partial \gamma}}\\=\gamma\frac{\partial\Re{\phi_{\alpha\beta}}}{\partial \omega},
\end{split}
\end{equation}
where we used the Cauchy-Riemann theorem and the fact that the phase $\phi_{\alpha\beta}(\omega)$ is real for $\omega$ being real. A similar approach has been used before to study lossy chaotic cavities\cite{Beenakker:Distributionreflection:01}.

The derivative of $\phi_{\alpha\beta}$ with respect to $\omega$ appearing in Eq.~\eqref{dphidw} has dimension of time, and we identify this time with the transmission time\cite{Landauer:Barrierinteraction:94} through the system $\partial\phi_{\alpha\beta}/\partial \omega=\tau_{\alpha\beta}$. Hence, we can relate the nonreciprocity parameter for weak dissipation to the transmission time in the following way
\begin{equation}
\begin{split}
    \Delta G\approx \abs{a(\omega+i\gamma)}^2(e^{-2\gamma\tau_{LR}}-e^{-2\gamma\tau_{RL}})\\\approx-2\gamma\abs{a(\omega)}^2(\tau_{LR}-\tau_{RL})
    \label{Delta_G-eq}
\end{split},
\end{equation}
showing that nonreciprocity caused by a featureless lead is related to the \textit{difference in transmission time} between the left-moving and right-moving states.

The transmission time $\tau_{\alpha\beta}$ is generally thought of as the time an electron spends inside the system, provided that it enters the system through lead $\beta$ and leaves through lead $\alpha$. It is different from the dwell time, which characterizes how long an electron stays inside the system regardless of where it starts or ends\cite{Winful:Tunnelingtime:06}. While these timescales can be thought of as actual transport timescales, especially in the case of ballistic transport, the so-called Hartman effect can be observed in tunneling, where the transmission time is independent of the tunnel barrier width, and as such the associated velocity can be arbitrarily high\cite{Hartman:TunnelingWave:62}. One thus needs to be careful about how these timescales are interpreted.

\subsection{Symmetry requirements}
We can get more insight into the nonreciprocity by studying how it depends on various physical symmetries of the system. First of all, we see that inversion simply corresponds to swapping the labels $R\leftrightarrow L$, and thus a system with inversion symmetry must have $t_{LR}=t_{RL}$, and thus no nonreciprocity.

We can also consider a system with time reversal symmetry. This corresponds to the $S$-matrix being symmetric, such that $S^T=S$. In this case, it is easy to see that $t_{LR}=t_{RL}$, and that there is no nonreciprocity.

Breaking inversion and time reversal symmetry is necessary for nonreciprocal conductance, but it is not sufficient. To see this, we can consider a multi-channel system. In such a system, $t_{LR}$ is a matrix, and the conductance is given by $\Tr[t_{LR}^\dagger t_{LR}]$. Inversion symmetry still implies that $t_{LR}=t_{RL}$, while time reversal symmetry now implies $t_{LR}=t_{RL}^T$, and both of these lead to the two nonlocal conductances being identical. However, we can also consider a broader class of symmetries, where $t_{LR}=U^\dagger t_{RL}^{(T)} U$ where $U$ is some unitary transformation which mixes the conduction channels of the system. If the system obeys such a symmetry, the conductance will still be reciprocal, so all such symmetries must be broken. Symmetries of this form include time-reversal or inversion, times spin rotation or gauge transformations.

\section{Rashba nanowire}

In this section, we consider an illustrative example of the above general consideration. A simple way to realize different transmission times in the two directions is a ballistic system where the Fermi velocity in different directions is different, such that the transmission time is $\tau_{\alpha\beta}\propto 1/{v_{\alpha\beta}}$ where $v_{\alpha\beta}$ is the Fermi velocity of particles moving from $\beta$ to $\alpha$.

\begin{figure}[t]
    \centering
    \includegraphics[width=\linewidth]{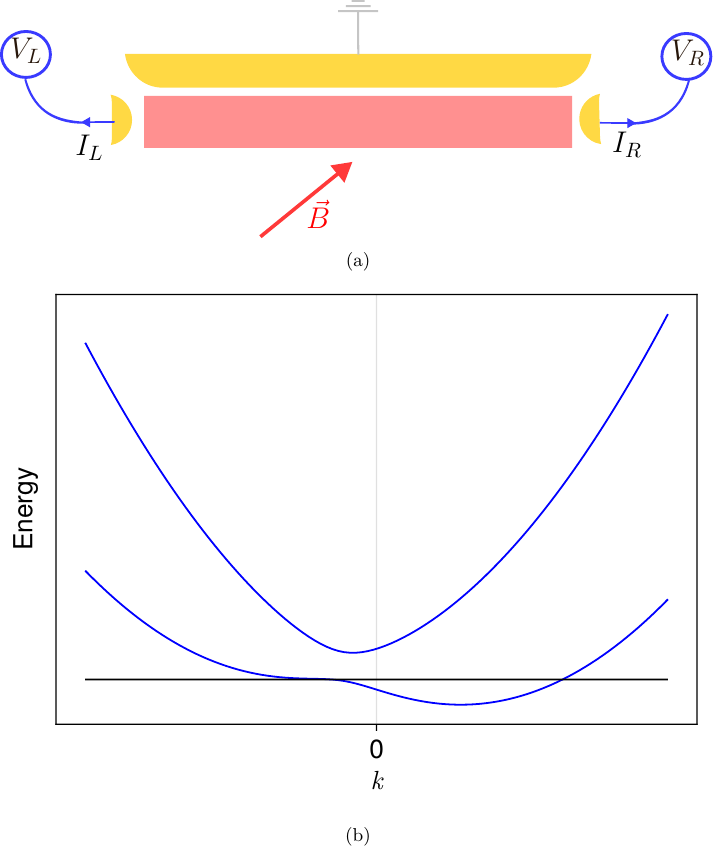}
    \caption{(a)A schematic of the nanowire in red connected to the leads in yellow. The magnetic field $\vec{B}$ is at some skewed angle compared to the spin-orbit direction. (b) Band structure of Rashba nanowire with skewed magnetic field, with the Fermi level marked in black. The Fermi velocity is given by the slope of the dispersion at the Fermi level. This slope is significantly larger for the right-movers than for the left-movers.}
    \label{fig:schematics}
\end{figure}

Such a system can be created by taking a nanowire with Rashba spin-orbit coupling and applying a Zeeman field which is skewed compared to the spin-orbit axis\cite{Zou:Nonreciprocalballistic:24}. We can then couple this system weakly to a lead along the length of the wire, such that the coupling is uniform in position and spin, as shown in Fig.~\ref{fig:schematics}(a). The Hamiltonian of this system is
\begin{equation}
    H=\frac{p^2}{2m}-\mu+\alpha\sigma_zp+\vec{B}\cdot \vec{\sigma}-i\gamma.
    \label{Hamiltonian}
\end{equation}
where $m$ is the effective electron mass in the nanowire, $\alpha$ is the spin-orbit strength, $\mu$ is the chemical potential, $\vec{B}$ is the applied Zeeman field, and $\gamma$ is the loss rate due to the coupling to the third (featureless) lead. To get a skewed Zeeman field, we choose $\vec{B}=(B\sin\theta,0,B\cos\theta)^T$ and choose $\theta\neq \frac{n\pi}{2}$ with integer $n$. This choice of $\theta$ ensures that both time-reversal and inversion symmetry is broken, and that neither can be restored by adding a simple spin rotation.

\begin{figure*}
    \centering
    \includegraphics[width=\textwidth]{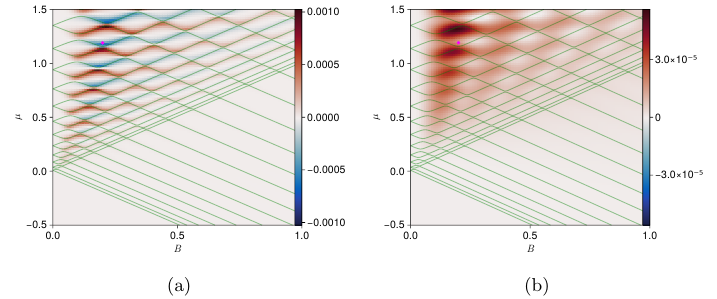}
    \caption{The nonreciprocity parameter $\Delta G$ as a function of chemical potential and Zeeman field for a dissipative Rashba nanowire of length $L=101$ at two different dissipation strengths. The green lines are the eigenenergies of Eq.(\ref{Hamiltonian}), and the magenta dot is the parameters for Figure \ref{fig:eta-gamma}. We note that there is both positive and negative nonreciprocity present for small dissipation strengths, while for large strengths there is only positive nonreciprocity. Parameters are $m=0.05$, $\alpha=0.2$, $\theta=\frac{\pi}{4}$ and (a)$\gamma=0.02$ or (b)$\gamma=0.1$.}
    \label{fig:etamap}
\end{figure*}

We can now look at the dispersion relation of this wire, as shown in Fig.~ \ref{fig:schematics}(b). We see that at the Fermi level, the left and right branches have different slopes, and thus different velocities. We thus expect a nonreciprocal response.
To look at this in more detail, we calculate the conductance of the device using Quantica\cite{San-Jose:pablosanjoseQuanticajl:25}. Figure \ref{fig:etamap} shows $\Delta G$ as a function of $B$ and $\mu$ for a particular set of parameters. We see that there is a nonreciprocal response, which is largest for relatively small but non-zero magnetic fields. This matches what we would expect from the dispersion relation of the wire, as the difference in the Fermi velocities is most pronounced for small but non-zero fields.

For small dissipation strengths, the nonreciprocity changes sign as the parameters and chemical potential are changed. While this might seem strange at first, it is easily explained by considering the fact that the upper band shows the opposite Fermi velocity difference, and that this upper band is also accessed for large enough chemical potentials. This also explains why the negative nonreciprocity is negligible compared to the positive at small $\mu$, but becomes more notable as $\mu$ is increased. We will return to why this effect depends on the dissipation strength in section \ref{fit}.

\section{Interference effect}
When looking at Fig.~\ref{fig:etamap}, one sees that the nonreciprocal response is largest near state anticrossings which implies an interference effect. To understand this, we consider the relation between the scattering matrix and the Green function\cite{Mahaux:SHELLMODELAPPROACH:69}
\begin{equation}
    S=1-W^\dagger G W,
\end{equation}
where $W$ is a vector containing the couplings between the system and the leads and $G$ is the retarded Green function.

Since the Green function also includes self-energy contributions from the left and right lead which are generally non-Hermitian, we write it in the biorthogonal form\cite{Brody:Biorthogonalquantum:13}
\begin{equation}
    G=\sum_n\frac{\ket{\psi_n^r}\bra{\psi_n^l}}{\omega-E_n}=\sum_n\frac{\rho_{\psi n}}{\omega-E_n+i\gamma}
    \label{GF}.
\end{equation}
where $E_n$ is generically some complex eigenenergy, and we have introduced the biorthogonal density matrix $\rho_{\psi n}=\ket{\psi_n^r}\bra{\psi_n^l}$.

We start out by considering the case where the transport is dominated by a single state, $\psi$. In that case, only one term in Eq.~\eqref{GF} contributes and the transmission can be written as
\begin{equation}
    t_{\alpha\beta}\approx\sqrt{\Gamma_\alpha\Gamma_\beta}\frac{\bra{x_\alpha}\rho_\psi\ket{x_\beta}}{\omega-E_\psi+i\gamma},
\end{equation}
where $\Gamma_{\alpha/\beta}$ are the coupling strengths to the $\alpha/\beta$ leads.

Since the dissipation difference is given by the difference in transmission times for the two-terminal system with $\gamma=0$, we can look at that case. Here, $\abs{t_{LR}}^2=\abs{t_{RL}}^2$ and hence $\bra{x_L}\rho_{\psi}\ket{x_R}=e^{i\delta}\bra{x_R}\rho_{\psi}\ket{x_L}$. If we now calculate the transmission times, we get that
\begin{equation}
    \tau_{LR}=\Im{\frac{1}{t_{LR}}\frac{d}{d\omega}t_{LR}}=\Im{\frac{e^{-i\delta}}{t_{RL}}\frac{d}{d\omega}t_{RL}e^{i\delta}}=\tau_{RL}
\end{equation}
where in the last step we use the fact that the phase difference between $t_{LR}$ and $t_{RL}$ is just given by $\delta$, which is energy-independent. We thus see that the transmission times in the two directions are the same.

We can instead consider a situation where two states contribute to the transport. In this case, we can write
\begin{equation}
\begin{split}
    t_{\alpha\beta}=\sqrt{\Gamma_\alpha\Gamma_\beta}\left(\frac{\bra{x_\alpha}\rho_{\psi n}\ket{x_\beta}}{\omega-E_n}+\frac{\bra{x_\alpha}\rho_{\psi m}\ket{x_\beta}}{\omega-E_m}\right)\\=t_{\alpha\beta}^n+t_{\alpha\beta}^m
\end{split}
\end{equation}
We see that $t_{LR}$ and $t_{RL}$ are related by a phase, but that this phase may be energy-dependent. Since the transmission time difference is given by the difference in phase between $t_{LR}$ and $t_{RL}$, we see that in this case we can have a transmission time difference.

\subsection{Higher-order effects}
\label{fit}
When looking at $\Delta G$ at different dissipation strengths, we see that the sign of the nonreciprocity can change at low dissipations, whereas it is always positive at higher dissipation. To understand this, we consider how $\Delta G$ depends on $\gamma$. In particular, if we assume the minimal model from above where the transport is primarily through two states, we can write
\begin{equation}
\begin{split}
    \Delta G\approx\Gamma_L\Gamma_R\abs{\frac{\bra{x_L}\rho_{\psi n}\ket{x_R}}{\omega-E_n+i\gamma}+\frac{\bra{x_L}\rho_{\psi m}\ket{x_R}}{\omega-E_m+i\gamma}}^2\\
    -\Gamma_L\Gamma_R\abs{\frac{\bra{x_R}\rho_{\psi n}\ket{x_L}}{\omega-E_n+i\gamma}+\frac{\bra{x_R}\rho_{\psi m}\ket{x_L}}{\omega-E_m+i\gamma}}^2
\end{split}
\end{equation}
which can be simplified to
\begin{widetext}
\begin{equation}
    \Delta G\approx\Gamma_L\Gamma_R\Re{\frac{\bra{x_L}\rho_{\psi n}\ket{x_R}\bra{x_R}\rho_{\psi m}^\dagger\ket{x_L}-\bra{x_R}\rho_{\psi n}\ket{x_L}\bra{x_L}\rho_{\psi m}^\dagger\ket{x_R}}{(\omega-E_n+i\gamma)(\omega-E^*_m-i\gamma)} }.
\end{equation}
\end{widetext}
Since we are interested in the $\gamma$-dependence of this expression, we simplify the $\gamma$-independent terms to get
\begin{equation}
    \Delta G\approx\Re{\frac{c}{(a-i\gamma)(b+i\gamma)} }.
\end{equation}
where $a$, $b$ and $c$ are complex numbers which are not dependent on $\gamma$. We can now use the fact that $\Delta G|_{\gamma=0}=0$ to see that $c=i r\cdot ab$ where $r\in \mathbb{R}$, and we thus have the form of the $\gamma$-dependence of $\Delta G$ at a single point in parameter space
\begin{equation}
\begin{split}
    \Delta G\approx\Re{\frac{ir\cdot ab}{(a-i\gamma)(b+i\gamma)} }\\=\frac{\Re{ir\cdot ab(a^*+i\gamma)(b^*-i\gamma)}}{\abs{a-i\gamma}^2\abs{b+i\gamma}^2} 
    \label{etaform}
\end{split}
\end{equation}
For relatively small $\gamma$, the numerator is dominated by terms proportional to $\mathrm{Re} [\abs{a}^2b-\abs{b}^2a]\gamma$, whereas for large $\gamma$ the dominant term is $\mathrm{Im}[ab]\gamma^2$. These terms can have different signs, which lead to a sign change as $\gamma$ increases.

\begin{figure}
    \includegraphics[width=\linewidth]{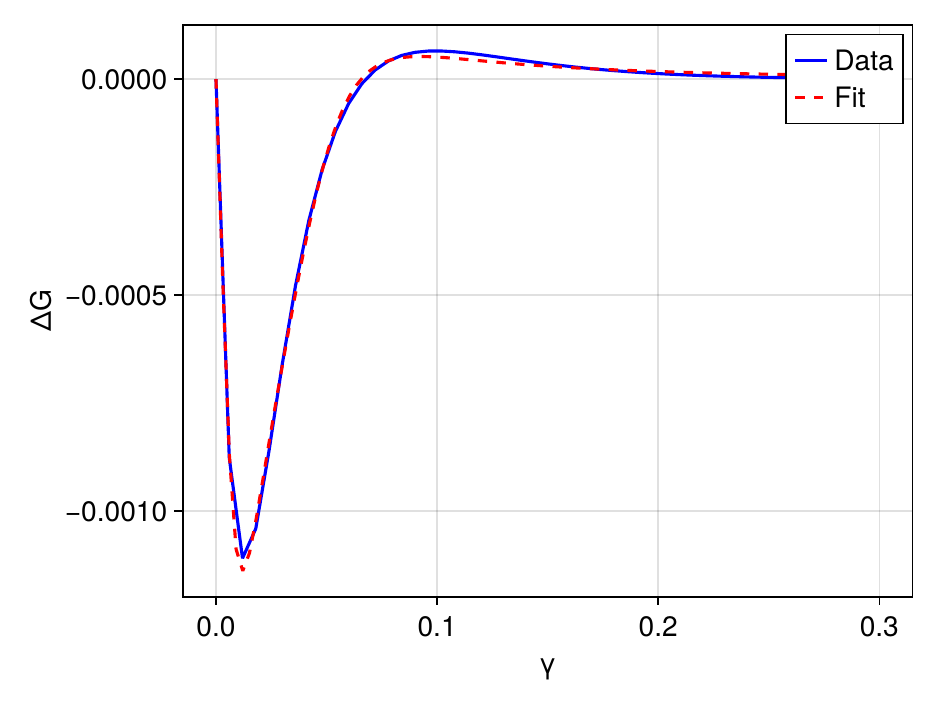}
    \caption{$\Delta G$ as a function of $\gamma$. The blue line is the numerical results, the dashed line is from a fit to (\ref{etaform}). Parameters are as in Fig.~\ref{fig:etamap}, and $B=0.2$, $\omega=1.2$.}
    \label{fig:eta-gamma}
\end{figure}
In Fig. \ref{fig:eta-gamma}, we calculate $\Delta G$ at a particular set of parameters as a function of $\gamma$, and then fit Eq.~(\ref{etaform}). This gives a fairly good, but not perfect, fit. This is expected since multiple states are relevant once the broadening induced by $\gamma$ is larger than the splitting of the states. These higher-order effects can also be understood as arising from the curvature of the band structure, which becomes relevant for increasing dissipation.

\section{Discussion}
We have shown that trivially non-Hermitian Hamiltonians can exhibit nonreciprocal transport, and described how this happens due to different transmission times in different directions. We have then considered an example system consisting of a Rashba nanowire with a skewed magnetic field, and shown that the different transmission times can be understood either from the band structure of the wire, or from the interference between the states within the finite wire. Finally, we have shown how higher-order effects in $\gamma$ modify the nonreciprocity.

We note that the effect described here is indeed measurable. Using numbers relevant for indium arsenide nanowires, we calculate conductances of $10^{-3}\sim 10^{-4} \frac{e^2}{h}$, which are well within what is measurable.

The results of this paper have implications for two fields of research. The first is the field of non-Hermitian physics in mesoscopic devices. Here, the fact that a nonreciprocal conductance can occur without the NHSE makes it more difficult to argue that the NHSE is actually present in various experimental proposals \cite{Paya:NonHermitianskin:26,Geng:Nonreciprocalcharge:23}. This gives rise to a need for more robust protocols.

The second field of research is simulation of mesoscopic devices, where a non-Hermitian term is often used as a phenomenological way to model dephasing. In this case, it is already known that trivially non-Hermitian terms can have effects on the calculated topological invariants\cite{ArayaDay:Identifyingbiases:25}. Our results show that even simpler observables like the conductance can be modified qualitatively by the presence of trivially non-Hermitian terms, including in relatively simple systems. This shows the importance of considering the physical nature of these terms and of treating them carefully.

{\it Data availability} The code to perform the calculations and make the plots is available at \cite{Solow:Nonreciprocalconductance:26}.

{\it Acknowledgements. } The authors are grateful to Piet Brouwer for useful discussions. This research was funded in part by the European Research Council (Grant Agreement No. 856526), the DFG Collaborative Research Center (CRC) 183 Project No. 277101999, the Knut and Alice Wallenberg Foundation (2023.0256) and the Göran Gustafsson Foundation for Research in Natural Sciences and Medicine.
\bibliography{My_Library}

\end{document}